\documentstyle[epsf]{elsart}
\begin{document}
\begin{frontmatter}
\title{Fixed-Point Hamiltonians in Quantum Mechanics}
\author[ITA]{T. Frederico}$^{,}$\footnote{tobias@fis.ita.cta.br,
$^2$delfino@if.uff.br,$^3$tomio@ift.unesp.br},
\author[UFF]{A. Delfino}$^{, 2}$,
\author[IFT]{Lauro Tomio}$^{, 3}$, 
\author[ITA]{V.S. Tim\'oteo}
\address[ITA] {Departamento de F\'\i sica,
Instituto Tecnol\'ogico de Aeron\'autica, CTA, 12228-900, 
S\~ao Jos\'e dos Campos, SP, Brasil}
\address[UFF] {Instituto de F\'\i sica, Universidade Federal Fluminense,
24210-900 Niter\'oi, RJ, Brasil}
\address[IFT]{Instituto de F\'\i sica Te\'orica, Universidade
Estadual Paulista,  01405-900, S\~{a}o Paulo, SP, Brasil} 
\date{\today }
\maketitle
\begin{abstract}
We show how to derive fixed-point Hamiltonians in quantum mechanics
from a proposed renormalization group invariance approach that
relies in a subtraction procedure at a given energy scale.
The scheme is valid for arbitrary interactions that originally contain
point-like singularities (as the Dirac-delta and/or its derivatives).
One example of diagonalization of a fixed-point Hamiltonian
illustrates the procedure for an interaction that contains regular and
singular parts. In another example, the fixed point Hamiltonian is
derived for the case of higher order singularities in the interaction.
\end{abstract}
\begin{keyword}
Renormalization, Renormalization group, Hamiltonian approach, 
scattering theory
\end{keyword}
\end{frontmatter}

Recently, it has increased the interest in the application of effective
theories to represent a more fundamental theory, as  Quantum
ChromoDynamics (QCD), which is too much complex to be solved exactly.
The program followed by Wilson and collaborators~\cite{wilglazek,wilperry} 
is of particular interest in this aspect, as it allows one to parametrize
the physics of the high momentum states and work with effective degrees of
freedom.  In ref.~\cite{wilglazek}, it was suggested the idea to use an
effective renormalized Hamiltonian that, in the interaction between
low-momentum states, includes the coupling with high momentum states.
The renormalized Hamiltonian carries the physical information
contained in the quantum system in states of high momentum. 
(The nonperturbative renormalization of 
refs.~\cite{LC,BPP}, for the light-front field theory, 
can also be related with the above procedure). 
As an example, in the nuclear physics context, the use of effective
interactions containing singularities at short distances is motivated by
the development of a chirally symmetric nucleon-nucleon interaction,
which contains contact interactions (Dirac-delta and its higher order
derivatives)~\cite{wein}.  Singular contact interactions have
also been considered in specific treatments of scaling limits 
and correlations between low-energy observables of three-body systems, in 
atomic and nuclear physics~\cite{amorim}. 
For other examples, see ref.~\cite{lepage}. 

A general non-perturbative renormalization scheme to treat 
singular interactions in  quantum mechanics has been proposed 
in ref.~\cite{rg00}, with applications in a nuclear physics example 
in ref.~\cite{fred99} and in an effective QCD-inspired theory 
of mesons in ref. ~\cite{pauli}. 
It generalizes some ideas suggested in refs.~\cite{fred95a}, by performing
$n$ subtractions in the free propagator of the singular scattering equation 
at an arbitrary energy scale. 
$n$ is the smallest number necessary to regularize the integral equation.  
The unknown short range physics related to the divergent part of the
interaction, are replaced by the renormalized strengths of the 
interaction, that are known from the scattering amplitude at some 
reference energy. 
In this context, the renormalization scale is given by an arbitrary 
subtraction point. And a sensible theory of singular interactions exists 
if and only if the subtraction point slides without affecting the 
physics of the renormalized theory \cite{wein1}. 

The subtraction point is the scale at which the quantum mechanical scattering 
amplitude is known~\cite{rg00}. 
A fixed-point Hamiltonian~\cite{wil,fisher,zinnjustin} should have the property 
to be stationary in the parametric space of Hamiltonians, as a function of the 
subtraction point~\cite{zinnjustin}. 
This property is realized through the vanishing derivative of the
renormalized Hamiltonian, in respect to the renormalization scale. 
This implies in the independence of the T-matrix in respect to
the arbitrary subtraction scale, and in the renormalization group
equations for the scattering amplitude. 
As shown in ref.~\cite{rg00}, the driving term of the $n-$subtracted
scattering equation changes as the subtraction point moves due to
the requirement that the physics of the theory remains unchanged. 
The driving term satisfies a quantum mechanical Callan-Symanzik (CS)
equation~\cite{CS}, i.e., a first order differential equation in respect
to the renormalization scale.  The renormalization group equation (RGE)
matches the quantum mechanical theory at scales $\mu$ and $\mu+d\mu$,
without changing its physical content~\cite{georgi}.

The purpose of the present letter is to show how one should obtain the 
renormalized (fixed-point) Hamiltonian for a quantum mechanical system consistent
with ref.~\cite{rg00}.
The general concept of fixed-point Hamiltonians is unified to the practical 
and useful theory of renormalized scattering equations~\cite{rg00}. 
The response to the relevant question on the existence and formulation of the
corresponding renormalized Hamiltonian  of quantum mechanical systems 
when the original interaction contains singular terms is given. This is important 
from the theoretical, as well as from the practical, point of view. 
The method is illustrated by the diagonalization of a
renormalized Hamiltonian and also by constructing the fixed-point Hamiltonian 
in a situation of higher singularities in the interaction.

We start by defining the renormalized
(fixed-point) Hamiltonian with the corresponding effective interaction
$V_{\cal R}$:
\begin{equation}
H_{\cal R} = H_0 + V_{\cal R},
\label{1}\end{equation}
where $H_0$ is the free Hamiltonian.
$H_{\cal R}$ and $V_{\cal R}$ are understood as `fixed-point' operators,
as they should not depend on the subtraction point.
The corresponding T-matrix equation of the above fixed-point Hamiltonian is 
given by the Lippmann-Schwinger
equation for the interaction $V_{\cal R}$,
\begin{eqnarray}
T_{\cal R}(E)=V_{{\cal R}}+V_{{\cal R}}G^{(+)}_0(E)T_{\cal R}(E) \ .
\label{2}
\end{eqnarray}

From the expression of $V_{\cal R}$, we should be able to derive the
$n-$subtracted T-matrix equation, the perturbative renormalization of the
T-matrix, and the CS equation satisfied by the driving term of the
subtracted form of the scattering equation.
So, considering the renormalization approach given in ref.~\cite{rg00},
the expression of the fixed-point interaction $V_{\cal R}$ is given
in terms of the driving operator of the $n-$th order subtracted equation.
For an arbitrary energy $E$, the driving operator is given by
$V^{(n)}(-\mu^2;E)$, where $-\mu^2$ is the subtraction point in the
energy space, that for convenience is chosen to be negative.  
$n$ is the number (order) of subtractions necessary for a complete 
regularization of the theory. 
Recalling from ref.~\cite{rg00}, the $n-$th order subtracted equation for
the T-matrix and also for the driving term  
$V^{(n)}\equiv V^{(n)}(-\mu^2;E)$ are, respectively, 
\begin{eqnarray}
&&T(E)=V^{(n)} (-\mu^2;E)
+ V^{(n)}(-\mu^2;E)G^{(+)}_n(E;-\mu^2)T(E) ,
\label{3}\\ 
{\rm and}\;\;\;\;
&&V^{(n)}\equiv \left[1-(-\mu^2-E)^{n-1}V^{(n-1)}
G^{n}_0(-\mu^2)\right]^{-1}V^{(n-1)}
+V_{sing}^{(n)}(-\mu^2) , \label{4}\\
{\rm where}\;\; 
&&G^{(+)}_n(E;-\mu^2)\equiv\left[(-\mu^2-E)G_0(-\mu^2)\right]^n 
G^{(+)}_0(E)
.\label{5} \end{eqnarray}
$G^{(+)}_0(E)=(E+{\rm i}\epsilon -H_0)^{-1}$ is the free Green's
function corresponding to forward propagation in time.
The higher-order singularities of the two-body potential are introduced 
in the driving term of the $n-$th order subtracted T-matrix 
through $V_{sing}^{(n)}(-\mu^2)$. The 
interaction, $V_{sing}^{(n)}(-\mu^2)$, are determined by physical
observables.

To obtain an expression for the fixed-point interaction 
$V_{\cal R}$, we first move the second term in the rhs of  
eq.~(\ref{3}) to the lhs; next, we add in both sides 
a term containing the driving term multiplied by the free Green's
function and the T-matrix:
\begin{eqnarray}
&&\left[1 + V^{(n)}
\left(G^{(+)}_0(E)-G^{(+)}_n(E;-\mu^2)\right) 
\right] T(E)=
V^{(n)} + V^{(n)}G^{(+)}_0(E) T(E). 
\label{6}
\end{eqnarray}
As in eq.~(\ref{4}), in order to simplify the notation, we drop
the explicit dependence of $V^{(n)}$ on $\mu^2$ and $E$.
Multiplying the above equation by the inverse of the
operator that is inside the square-brackets, we can identify the
fixed-point interaction $V_{\cal R}$ considering the equivalence
we want to establish between eqs.~(\ref{2}) and (\ref{3}).
So, in order to have $T_{\cal R}(E) = T(E)$, 
\begin{eqnarray}
V_{{\cal R}} =\left[{1+V^{(n)}\left(G^{(+)}_0(E)-
G^{(+)}_n(E;-\mu^2)\right)}\right]^{-1} 
V^{(n)} \   .\label{7}
\end{eqnarray}
By doing this, we are identifying the fixed-point Hamiltonian,
eq.~(\ref{1}) in our renormalization scheme.

One could observe that the above fixed-point interaction is not
well defined for singular interactions; nevertheless, the corresponding
T-matrix is finite, as we have shown from the equivalence with
the $n-$th order subtracted equation for the T-matrix, given in
ref.~\cite{rg00}. Essentially, the renormalization subtraction procedure
that was used in ref.~\cite{rg00} was instrumental to write the 
renormalized fixed-point interaction given in eq.~(\ref{7}).
In the following, considering the above equivalence, it should be
understood that our T-matrix refers to the renormalized one
($T=T_{\cal R}$), such that we drop the index $\cal R$ from it.
However, there is a clear distinction between $V^{(n)}$ and $V_{\cal R}$,
such that the index cannot be dropped in this case.

In a concrete numerical application, when working with the operator 
$V_{{\cal R}}$, we should use a momentum regulator, that 
could be, for example, 
a sharp ultraviolet momentum cutoff ($\Lambda $).
By performing the limit $\Lambda\rightarrow \infty$, the results 
should be the same as the ones obtained through the direct use of the
subtracted scattering equations.
In particular, it  implies that the eigenvalues of a renormalized 
Hamiltonian are stable in the limit $\Lambda\rightarrow \infty$, and
agree with the correct results obtained from the subtracted integral 
equations. This fact will be illustrated.

The physical informations that apparently are lost due to the
subtractions made in the intermediate free propagator are kept in
$V^{(n)}$. In the case of the two-body system, it contains the
renormalized coupling constants of the singular part of the interaction,
given at some energy scale $-{\mu}^2$. 
We should also notice that one subtraction in the kernel of
the integral equation is enough to obtain meaningful physical results
from a T-matrix that contains a Dirac-delta interaction; 
and at least three subtractions are necessary if the singularity
of the interaction corresponds to the Laplacian of the
Dirac-delta function~\cite{rg00}.
In the case of a non-singular short-range interaction ($V$) for which 
$V_{sing}^{(n)}=0$, we immediately identify that $V_{{\cal R}}= V $,
which is not surprising since the renormalized T-matrix is the T-matrix
obtained from the conventional Lippmann-Schwinger equation.

We have initially defined a nonperturbative renormalized
interaction $V_{\cal R}$, although it is natural to ask about the
perturbative expansion of the  renormalized T-matrix in powers of
$V^{(n)}$. One can easily show that, the renormalized interaction 
contains the necessary counterterms to include the subtractions in each 
perturbative order. 
This can be answered by performing the expansion of the
renormalized interaction $V_{\cal R}$, eq.~(\ref{7}), in powers of
$V^{(n)}$ of order $m$.
Consequently, for each order $m$ one can obtain the corresponding 
perturbative expansion:
\begin{eqnarray}
T^{(m)}(E)=V^{(n)} \sum_{k=0}^{m-1} \left[ G^{(+)}_n(E;-\mu^2)
V^{(n)} \right]^k \ .  \label{10}
\end{eqnarray}
Thus, each intermediate state propagation up to order $m$ is subtracted at
the energy scale $-\mu^2$. This shows that the fixed-point Hamiltonian
includes the necessary counterterms in each perturbative order, which
impose a subtraction of each intermediate propagation in the 
perturbative expansion of the scattering amplitude. As a simple remark,
eq.~(\ref{10}) could be obtained from eq.~(\ref{3}), by iterating
$m-1$ times and truncating at the order $m$.

The subtraction point is arbitrary in the definition of the renormalized
interaction and in principle it can be moved~\cite{wein1}. The change in
the subtraction point requires the knowledge of the driving term $V^{(n)}$
at the new energy scale. The coefficients that appear in the driving 
term $V^{(n)}$ come from the prescription used to define them.
The renormalization group method can be used to arbitrarily
change this prescription constrained by the invariance of the
physics under dislocations of the subtraction point. This condition 
demands the renormalized potential $V_{{\cal R}}$ to be
independent on the subtraction point. 
It gives a definite prescription to modify $V^{(n)}$ in eq.~(\ref{3}), 
without altering the predictions of the theory. 
So, 
\begin{eqnarray}
\frac{\partial V_{\cal R}}{\partial \mu^2}= 0 
\;\;\;\; {\rm and}\;\;\;\; 
\frac{\partial H_{\cal R}}{\partial \mu^2}= 0 ;
\label{11}\end{eqnarray}
the renormalized Hamiltonian does not depend on 
$\mu$; it is a fixed-point Hamiltonian in this respect.
From eqs.~(\ref{2}) and (\ref{11}), the renormalized 
T-matrix also does not dependent on $\mu$,
\begin{eqnarray}
 \frac{\partial T(E)}{\partial \mu^2} =0 \ .
\label{12}
\end{eqnarray}
The renormalization group equation, satisfied by the running
driving term $V^{(n)}$ in respect to the sliding subtraction point, 
follows by using the null-derivative condition,
given by eq.~(\ref{11}) in eq.~(\ref{7}):
\begin{eqnarray}
 \frac{\partial V^{(n)}}{\partial \mu^2} = -V^{(n)}
\frac{\partial G^{(+)}_n(E;-{ \mu}^2) }{\partial \mu^2}
V^{(n)} \ .
\label{13}
\end{eqnarray}
Thus, we reobtain the nonrelativistic Callan-Symanzik
equation~\cite{rg00}, which reflects the invariance of the subtracted
scattering equation under the modification of the subtraction point. 
The equation (\ref{13}) expresses the invariance of the renormalized
T-matrix under dislocation of the subtraction point. The boundary
condition is given by $V^{(n)}= V^{(n)}(-{\overline{\mu }}^{2};E)$ at the 
reference scale $\overline{\mu }$. 
The subtracted potential $V^{(n)}$ obtained by solving the
differential equation (\ref{13}) is equal to the T-matrix for 
$E=-\mu^{2}$, relating the sliding scale to the energy dependence of the
T-matrix itself. In particular, for $n=1$, eq.~(\ref{13}) is the
differential form of the renormalized equation for the T-matrix
\begin{equation}
\left. \frac{d}{dE}T(E)\right| _{E=-\mu ^{2}}=-T(-\mu ^{2})G_{0}^{2}(-\mu
^{2})T(-\mu ^{2}).  \label{14}
\end{equation}
This concludes our discussion on the invariance of the fixed-point
Hamiltonian under renormalization group transformation, in case of
singular potentials. 
As we have seen, the fixed-point Hamiltonian contains the renormalized 
coefficients/operators that carry the physical informations of the
quantum mechanical system, as well as all the necessary counterterms that
make finite the scattering amplitude. 

Next, we present two examples to illustrate the present approach. 
In the first example, the numerical diagonalization of 
the regularized form of the fixed-point Hamiltonian for a two-body system 
with a Yukawa plus a Dirac-delta interaction is performed, 
and the observable (eigenvalues) are shown to be independent on the 
energy-point $-\mu^2$ where the subtraction is done; the results are 
shown to be stable in the infinite momentum cutoff. In another example, 
we derive the explicit form of the renormalized potential for an example 
of four-term singular bare interaction.

\noindent 1. {\bf Renormalized Hamiltonian diagonalization}

Let us consider a two-body system with a Yukawa plus Dirac-delta 
interaction
\begin{equation}
\langle \vec{p} |V_{\cal R}| \vec{q} \rangle 
= \langle \vec{p} |V| \vec{q} \rangle +\frac{\lambda_\delta}{2\pi^2}
=\frac{1}{2\pi^2}
\left(\frac{-2}{|\vec{p}-\vec{q}|^2+\eta^2}+\lambda_{\delta}\right),
\label{15}
\end{equation}
where $\lambda_\delta$ is the singular part of the renormalized or
fixed point interaction; and $\eta$ is a constant that is given by the 
inverse of the range of the regular part of the interaction.
The strength $\lambda_\delta$ can be derived from the full renormalized 
T-matrix given in eq.~(\ref{2}):
\begin{eqnarray}
\left(1-VG^{(+)}_0(E)\right)T_{\cal R}(E)= 
V+ |\chi\rangle\frac{ \lambda_{\delta}}{2\pi^2}\langle\chi| 
\left[1+ G^{(+)}_0(E)T_{\cal R}(E)\right] \ .
\label{18}
\end{eqnarray}
By defining the  T-matrix of the regular potential $V$ as
$T^V(E)= \left(1-VG^{(+)}_0(E)\right)^{-1}V$, and
using the identity
$\left(1-VG^{(+)}_0(E)\right)^{-1}
= \left(1+T^V(E)G^{(+)}_0(E)\right)$, 
one can easily obtain
\begin{eqnarray}\hskip -0.7cm
T_{\cal R}(E)&=&T^V(E)+ 
\frac{\left( 1+T^V(E)G^{(+)}_0(E)\right)|\chi\rangle
\langle\chi|\left(1+G^{(+)}_0(E)
T^V(E)\right)}
{\displaystyle \frac{2\pi^2}{\lambda_{\delta}}-
\langle\chi|G^{(+)}_0(E)|\chi\rangle -
\langle\chi|G^{(+)}_0(E)T^V(E)G^{(+)}_0(E)|\chi\rangle} 
\ . \label{19}
\end{eqnarray}
In this case, one subtraction is enough to render finite the theory.
At the subtraction point $-\mu^2$, the above equation  defines 
the T-matrix of eq. (\ref{4}) for $n=1$:
\begin{eqnarray}
V^{(1)}&=&T^V(-\mu^2)+ 
\frac{\left( 1+T^V(-\mu^2)G_0(-\mu^2)\right)|\chi\rangle
\langle\chi|\left(1+G_0(-\mu^2)T^V(-\mu^2)\right)}
{\displaystyle \frac{2\pi^2}{\lambda_{\delta}}-
\langle\chi|G_0(-\mu^2)|\chi\rangle -
\langle\chi|G_0(-\mu^2)T^V(-\mu^2)G_0(-\mu^2)|\chi\rangle} 
\ . \label{19a}
\end{eqnarray}
The denominator of the second term of eq.(\ref{19a}) at an
arbitrary energy $-\mu^2$ is defined by a constant $C^{-1}(-\mu^2)$, such that
\begin{equation}
2\pi^2C^{-1}(-\mu^2)\equiv
2\pi^2\lambda_{\delta}^{-1}-
\langle\chi|G_V(-\mu^2)|\chi\rangle \ , \label{16b}
\end{equation}
where  $G_V(-\mu^2)=-\left(\mu^2+H_0+V\right)^{-1}= 
G_0(-\mu^2)\left(1+T^V(-\mu^2)G_0(-\mu^2)\right) $.
The fixed-point structure of the renormalized potential (\ref{15}), 
$\partial V_{\cal R}/\partial\mu^2=0$ 
($\partial\lambda_\delta/\partial\mu^2=0$),
defines the functional form of $C(-\mu^2)$ on the subtraction point:
\begin{equation} 
2\pi^2C^{-1}(-\mu^{\prime 
2})-2\pi^2C^{-1}(-\mu^{2})=\langle\chi|\left(
G_{V}(-\mu^{2}) - G_{V}(-\mu^{\prime 2}) 
\right)|\chi\rangle .
\end{equation} 
In case that the subtraction point is chosen to be one of the binding
energies of the physical system, $\mu^2=\mu^2_B$, $C^{-1}(-\mu_B^2)=0$. 
So, from eq.(\ref{19a}), we obtain
\begin{equation}
\lambda_{\delta}^{-1}=
\frac{2}{\pi}\int_0^\Lambda dp p^2
\frac{-1}{\mu_B^2+p^2}
 + \frac{1}{2\pi^2}\int d^3p 
\frac{\Theta(\Lambda^2-p^2)}{\mu_B^2+p^2}
\int d^3q
\frac{\Theta(\Lambda^2-q^2)}{\mu_B^2+q^2}
{\langle \vec{p} |T^V(-\mu_B^2)| \vec{q} \rangle} ,
\label{16}
\end{equation}
where the momentum cutoff $\Lambda$ was included through the 
step function $\Theta(x)$ (defined as $0$ for $x<0$ and $1$ for $x>0$).
With the cutoff parameter $\Lambda$ assuming its exact limit (infinite),
$\lambda_{\delta}$ contains the divergences in the momentum integrals 
that exactly cancels the infinities in eq.(\ref{2}). 
We have to emphasize that, the role of the cutoff parameter 
$\Lambda$ is just of a regulator of the integrals that, at the end, should 
disappear in a simple limit $\infty$, without affecting the physical results.
The relevant scale parameter, where the physical information is supplied, is
the energy-point $-\mu^2_B$. As it will be shown, the results will not be
affected by a specific choice of the position of the subtraction point.

Once $V_R$ is defined, as in eq.(\ref{15}), we can perform the 
numerical diagonalization of the fixed-point Hamiltonian (\ref{1}),
and obtain the bound-states $\varepsilon$. 
The corresponding Schr\"odinger equation, 
in the $s-$wave, and in units such that $\hbar^2/(2m)=1$, can be written 
as 
\begin{eqnarray}
H_{\cal R}\Psi(p)= p^2\Psi(p)+ 
{\frac{2}{\pi}}\int_0^{\Lambda} q^2 dq 
\left[\frac{-1}{2pq}\ln\left(\frac
{\eta^2+(p+q)^2}{\eta^2+(p-q)^2}
\right)+\lambda_\delta\right]
\Psi(q) 
=\varepsilon\Psi(p) \ ,
\label{20}
\end{eqnarray}
where $\lambda_\delta$, given by the renormalization prescription,
keeps constant $\mu_B^2$, chosen as 
one of the bound-state energies.
To solve numerically this example, we first choose arbitrarily a regular 
{\it reference potential}, with the matrix elements given by 
$\langle\vec{p}|V|\vec{q}\rangle=$ $- \left[
\pi^2 \left(|\vec{p}-\vec{q}|^2 + \eta^2\right)\right]^{-1} $ 
$ - \left[\pi^2\left(|\vec{p}-\vec{q}|^2 + \eta^2_S\right)\right]^{-1} $, 
with $\eta^2=0.01$ and $\eta_S^2=1$ ($\eta^2$ and 
$\eta^2_S$, as well as the energies $\varepsilon$, are given in 
units of inverse-squared length). This reference potential produces 
three bound-state energies: $\varepsilon^{(0)}=-$2.3822, 
$\varepsilon^{(1)}=-$0.20297 and $\varepsilon^{(2)}=-$0.020643. 
Next, the short-range part of the reference potential is replaced by the 
Dirac-delta interaction.
This singular interaction together with the long-range part is 
renormalized with the 
physical condition being supplied by one of the bound-state energies. 
This is the observable that is supposed to be known in our hypothetical 
example. So, the subtraction point $-\mu^2$ will be given by this energy. 
At the same time that $\mu^2$ is regularizing the formalism, via a 
subtraction procedure, it is also carrying the relevant physical 
information in the present Hamiltonian renormalization approach.
\begin{figure}
\hskip -5cm .\vskip -3cm \hskip 5cm 
\begin{center}
{\setlength{\epsfxsize}{10cm}
\epsfbox{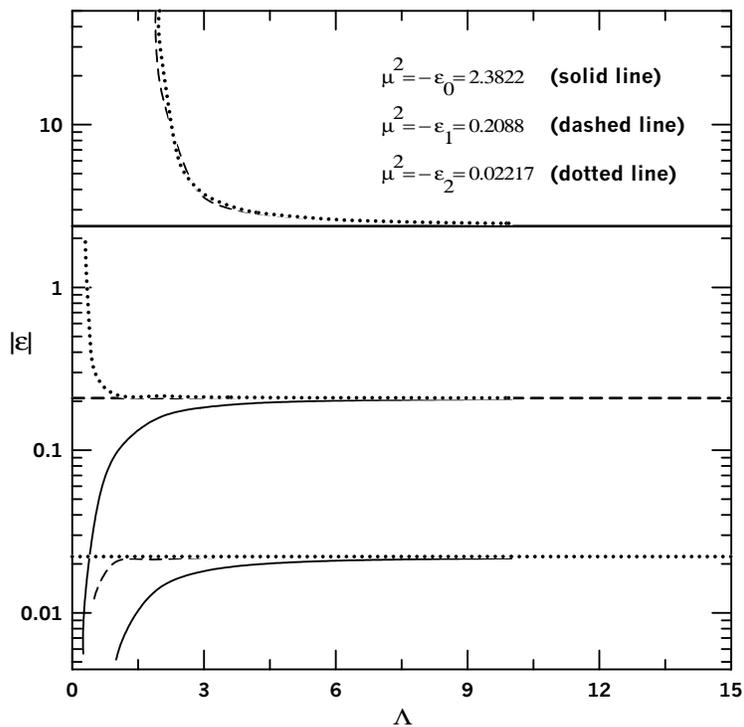}}
\end{center}
\caption
{The bound-state energies, $\varepsilon$, of the Hamiltonian given in 
example 1, are shown to converge to the same exact results, in the
limit $\Lambda\to\infty$ ($\Lambda=$ momentum cut-off), irrespectively
to the value of the energy-scale parameter $-\mu^2$ that is used to 
renormalize the theory. One of the three bound-state energies is used 
to define the subtraction point $-\mu^2$; the other two 
result from the diagonalization of the renormalized Hamiltonian.
$\mu$, $\Lambda$, and $\sqrt{|\varepsilon|}$ are given in units of 
momentum.
}
\end{figure}
\vskip 0.5cm

Our results are shown in Fig. 1 for the eigenvalues of the Hamiltonian,
as functions of the momentum cutoff parameter $\Lambda$. 
The specific choice of the subtraction point $-\mu^2$ (one of the three
straight lines) does not affect the final results.  
They are exact in the limit $\Lambda\to\infty$. 
Fig. 1 shows three sets of results; for each one, the value of $-\mu^2$
has a specific definition, given by one of the ``known" energies.
In solid lines, we have the results for the first and second
excited states, when $\mu^2=$ $-\varepsilon^{(0)}=$ 2.3822. The exact
results, $\varepsilon^{(1)}=-$0.2088 and $\varepsilon^{(2)}=-$0.02217,
are reached when $\Lambda\to\infty$. In the same way, we obtain the 
results given by the other two sets, with $\mu^2=$ $-\varepsilon^{(1)}=$ 
0.2088 (dashed-lines), and with $\mu^2=$ $-\varepsilon^{(2)}=$ 
0.02217 (dotted-lines). As shown, in this diagonalization procedure, the 
results are stable when $\Lambda\to\infty$, and converge to the 
exact values, that can be given by the real poles of the T-matrix: 
$\varepsilon^{(0)}=-$2.3822, $\varepsilon^{(1)}=-$0.2088 and 
$\varepsilon^{(2)}=-$0.02217.

This example gives a simple and clear picture about what we have stated in 
Eq.~(\ref{11}): that the renormalized Hamiltonian does not depend on the 
choice of $\mu$; it is a fixed-point Hamiltonian in this respect. \ The 
momentum cutoff $\Lambda$, used as an instrumental regulator, disappears 
in the present approach as a natural infinite limit of the integrals, 
where all the infinities presented in the formalism are canceled. 

\noindent 2. {\bf Renormalized Hamiltonian for a four-term singular
interaction}

Here we derive the explicit form of the renormalized potential
for an example of four-term singular bare interaction that, after
partial-wave decomposition to the $s-$wave, is given by

\begin{equation}
\langle p|V|q\rangle=\sum_{i,j=0}^{1}\lambda_{ij} p^{2i} q^{2j}
\;\;\;(\lambda_{ij}=\lambda_{ji}^*) \ .  \label{22}
\end{equation}

The renormalized strengths of the interaction are known from the physical
scattering amplitude at reference energy $-\overline\mu^2$
\begin{eqnarray}
\langle p| T(-{\overline \mu}^2)|q\rangle=\lambda_{{\cal R}00}
 +\lambda_{{\cal R}10} (p^2+q^2)+\lambda_{{\cal R}11} p^2 q^2 \ ,  
\label{23}
\end{eqnarray}
where for simplicity we suppose $\lambda_{{\cal R}10}$ real. The physics
 of the two-body system with the bare interaction of eq.~(\ref{22})
is completely defined by the values of the renormalized strengths,
$\lambda_{{\cal R}ij}$, at the reference energy $-\overline\mu^2$.

In the Lippman-Schwinger equation, the potential given by eq.~(\ref{22}) 
implies in integrals that diverge at most as $p^5$, requiring at least 
three subtractions to obtain finite integrals. With $n=3$ in 
eq.~(\ref{7}), from the recurrence relationship (\ref{4}), we obtain the 
following equations:
\begin{eqnarray}
&&\langle p|V^{(1)}(-{\overline\mu}^2)|q\rangle =\lambda_{{\cal R}00} \; ,
\;\;\;\;\; \langle p|V^{(2)}(-{\overline\mu}^2;k^2)|q\rangle =\left[
\lambda_{{\cal R}00}^{-1}+ I_0\right]^{-1} \; ,  \nonumber \\
&&\langle p|V^{(3)}(-{\overline\mu}^2;k^2)|q\rangle = \overline{\lambda}_{
{\cal R}00} +\lambda_{{\cal R}10}(p^2+q^2) +\lambda_{{\cal R}11} p^2q^2\ ,
\label{24} \\
{\rm with} 
&&\overline{\lambda}_{{\cal R}00}\equiv\left[\lambda_{{\cal 
R}00}^{-1} +I_0+I_1\right]^{-1}\nonumber\\
{\rm and}
&&I_{i=0,1}\equiv I_i(k^2,{\overline\mu}^2)\equiv
\frac{2}{\pi}\int^\infty_0dqq^2
\frac{({\overline\mu}^2+k^2)^{1+i}}{({\overline\mu}^2+q^2)^{2+i}} =
\frac{({\overline\mu}^2+k^2)^{1+i}} {(2{\overline\mu})^{1+2i}}\ 
\label{25}.\end{eqnarray}
Note that the singular term, as shown in eq.~(\ref{4}), is 
introduced for $n=3$ in $V^{(3)}$ of eq.~(\ref{24}).
Also, when $k^2=-\mu^2$ we have $I_i=0$ and
$\overline{\lambda}_{{\cal R}00} = \lambda_{{\cal R}00}$.
The renormalized interaction is obtained analytically in this example.
By introducing $V^{(3)}(-\overline{\mu }^{2},k^{2})$ 
of eq.~(\ref{24}) in eq.~(\ref{7}), 
\begin{eqnarray}
\langle p|V_{{\cal R}}|q\rangle =\sum_{i,j=0}^{1}\Lambda
_{ij}(k^{2})p^{2i}q^{2j}\;\;\;(\Lambda _{ij}=\Lambda _{ji})\ ,
\label{26}
\end{eqnarray}
where $\Lambda _{ij}$, that will not depend on the
subtraction point, are given by:
\begin{eqnarray}
\Lambda _{00}(k^{2})&=&
\frac
{\overline{\lambda }_{{\cal R}00} -
\left(\overline{\lambda}_{{\cal R}00} K_{1} +
      \lambda _{{\cal R}10}K_{2}\right)\Lambda _{10}}
{1+\overline{\lambda}_{{\cal R}00}K_{0}+\lambda_{{\cal R}10}K_1
} ;  \label{27} \\
\Lambda _{11}(k^{2})&=&
\frac
{\lambda_{{\cal R}11} -
\left(\lambda_{{\cal R}10} K_{0} +
      \lambda _{{\cal R}11}K_{1}\right)\Lambda _{10}}
{1+\lambda_{{\cal R}11}K_2+\lambda_{{\cal R}10}K_1 
}
;  \label{28} \\
\Lambda_{10}(k^{2})&=&
\frac{\lambda_{{\cal R}10}+
(\lambda_{{\cal R}10}^{2}-\overline{\lambda }_{{\cal R}00}
\lambda _{{\cal R}11})K_{1}} 
{1+D+(\lambda_{{\cal R}10}^2 - \overline{\lambda }_{{\cal R}00}
\lambda_{{\cal R}11})(K_1^2-K_0 K_2) 
},  \;\;\;\label{29}
\end{eqnarray}
with 
$D\equiv\overline{\lambda}_{{\cal R}00}K_{0}+\lambda_{{\cal R}10}K_1 
+\lambda_{{\cal R}11}K_2+\lambda_{{\cal R}10}K_1$ and 
\begin{eqnarray}
K_{i=0,1,2}&\equiv&K_{i}(k^{2},{\overline{\mu }}^{2})
\equiv \frac{2}{\pi}
\int_{0}^{\infty }\frac{dqq^{2i+2}}{k^2-q^2}
\left[1- \left(\frac{\overline{\mu}^2+k^2}
{\overline{\mu}^2+q^2}\right)^3\right]
.\label{31} \end{eqnarray}
$K_{i}$ are the divergent integrals that exactly cancels the
infinities of the Lippman-Schwinger equation obtained with the 
renormalized interaction, eq.~(\ref{26}).
The integrands of these integrals are given by the kernel of
eq.~(\ref{7}) with $n=3$.

The derivatives ${\partial \Lambda_{ij}}/{\partial\mu^2}$ vanish due
to the arbitrariness of the subtraction point. These conditions on the
derivatives are given by the explicit form of eq.~(\ref{11}) in the
case of the four-term singular potential. The dependence of the
coefficients $\lambda_{{\cal R}00}$, $\lambda_{{\cal R}10}$ and
$\lambda_{{\cal R}11}$ on the sliding scale $\mu$ can be computed either 
from the above conditions or directly from eq.~(\ref{13}). The boundary
condition of the RGE first order differential equation should be given for
each scattering energy $k^2$, considered as a parameter,
at the reference value of the subtraction point $-\overline\mu^2$.
This concludes our example of a derivation of the renormalized
Hamiltonian, with the corresponding renormalized interaction given by
(\ref{26}).

In summary, the fixed-point Hamiltonian emerges as a consequence 
of the renormalized $n$-subtracted scattering equation for the T-matrix. 
It does not depend on the 
position of the subtraction point, $-\mu^2$, where the physical 
information is supplied to the theory. As shown, it naturally includes the
renormalization group invariance properties of quantum mechanics 
with singular interactions, as expressed by the nonrelativistic 
Callan-Symanzik equation.  
Finally, we should emphasize the wide range of applicability of 
renormalized Hamiltonians, from atomic and nuclear physics 
models to effective theories of QCD (see, for example, refs. 
\cite{amorim}-\cite{pauli}). It would be of interest a comparison 
between the present nonperturbative Hamiltonian renormalization 
approach with other given formalisms; as, for example, the approach 
considered in ref.~\cite{birse}. 
However, a caution is necessary when doing such comparison, as
one should note that, in the present work, the invariance of the 
Hamiltonian is with respect to a subtraction energy scale, in the
limit of infinite momentum cutoff.
The present Hamiltonian renormalization approach is particularly 
useful when several discrete eigenvalues are possible, since it can be 
diagonalized, in a regularized form, in order to 
obtain physical observables that are well defined in the infinite 
cutoff limit. 

This work was partially supported by Funda\c c\~ao de Amparo \`a Pesquisa do
Estado de S\~ao Paulo (FAPESP) and Conselho Nacional de Desenvolvimento
Cient\'\i fico e Tecnol\'ogico (CNPq).

\end{document}